\documentclass[journal]{IEEEtran}
\usepackage{amsmath,amssymb,epsfig,cite}
\usepackage{latexsym,amssymb,amsmath,graphicx,epsf,cite,bbm,graphics}
\usepackage{setspace}

\newcommand{\defi}{\stackrel{\bigtriangleup}{=}}

\def\vw{{\vec{w}}}
\def\nn{{\nonumber}}

\def\vu{{\vec{u}}}
\def\vw{{\vec{w}}}

\def\cR{{\mathcal{R}}}

\newcommand{\vv}{\mbox{$\vec{v}$}}
\renewcommand{\vec}[1]{\mbox{\boldmath${#1}$}}
\renewcommand{\mathbf}[1]{\mbox{\boldmath${#1}$}}

\newcommand{\argmin}[1]{\underset{#1}{\operatorname{argmin\,}}}

\def\b0{{\mathbf{0}}}

\def\bb{{\mathbf{b}}}

\def\bff{{\mathbf{f}}}

\def\bs{{\vec{s}}}

\def\bu{{\mathbf{u}}}
\def\bv{{\mathbf{v}}}
\def\bw{{\mathbf{w}}}
\def\bx{{\mathbf{x}}}
\def\by{{\mathbf{y}}}

\def\bA{{\mathbf{A}}}
\def\bB{{\mathbf{B}}}
\def\bC{{\mathbf{C}}}
\def\bD{{\mathbf{D}}}
\def\bE{{\mathbf{E}}}

\def\bH{{\mathbf{H}}}
\def\bI{{\mathbf{I}}}

\def\bV{{\mathbf{V}}}

\newtheorem{remark}{Remark}

\newcommand{\bet}{\begin{table}}
\newcommand{\eet}{\end{table}}
\newcommand{\btt}{\begin{tabular}}
\newcommand{\ett}{\end{tabular}}
\newcommand{\bec}{\begin{center}}
\newcommand{\eec}{\end{center}}
\newcommand{\bef}{\begin{figure}}
\newcommand{\eef}{\end{figure}}
\newcommand{\beq}{\begin{eqnarray}}
\newcommand{\eeq}{\end{eqnarray}}
\newcommand{\beqs}{\vspace*{-0.05in}\begin{eqnarray}}
\newcommand{\eeqs}{\end{eqnarray}\vspace*{-0.05in}}
\newcommand{\bit}{\begin{itemize}}
\newcommand{\eit}{\end{itemize}}
\newcommand{\bed}{\begin{description}}
\newcommand{\eed}{\end{description}}
\newcommand{\ben}{\begin{enumerate}}
\newcommand{\een}{\end{enumerate}}
\newcommand{\bis}{\vspace*{-0.05in}\begin{itemize}\small}
\newcommand{\eis}{\end{itemize}\normalsize\vspace*{-0.05in}}

\newcommand{\eref}[1]{(\ref{#1})}
\newcommand{\fig}[1]{Fig.\ \ref{#1}}

\def\-{\! - \!}
\def\+{\! + \!}
\def\={\! = \!}
\def\>{\! > \!}

\def\ninept{\def\baselinestretch{0.98}}
\ninept
\begin{document}
\title{Low Complexity Turbo-Equalization: A Clustering Approach} \vspace{-0.6cm}
\author{Kyeongyeon Kim, Jun Won Choi, Suleyman S. Kozat, {\em Senior Member}, IEEE, and~Andrew~C.~Singer, {\em Fellow}, IEEE \thanks{K.Kim is with Samsung Electronics. J.W. Choi is with Qualcomm.  A.C. Singer is with the University of Illinois at Urbana-Champaign, Urbana IL 61801, email: acsinger@illinois.edu. S. S. Kozat is with the Competitive Signal Processing Laboratory at the Koc University, Istanbul, Turkey, email: skozat@ku.edu.tr, tel: 02123381684.}}\vspace{-0.6cm}
\maketitle
\begin{abstract}
We introduce a low complexity approach to iterative equalization and
decoding, or ``turbo equalization'', that uses clustered models to
better match the nonlinear relationship that exists between likelihood
information from a channel decoder and the symbol estimates that arise
in soft-input channel equalization.  The introduced clustered turbo
equalizer uses piecewise linear models to capture the nonlinear
dependency of the linear minimum mean square error (MMSE) symbol
estimate on the symbol likelihoods produced by the channel decoder and
maintains a computational complexity that is only linear in the
channel memory.  By partitioning the space of likelihood information
from the decoder, based on either hard or soft clustering, and using
locally-linear adaptive equalizers within each clustered region, the
performance gap between the linear MMSE equalizer and low-complexity,
LMS-based linear turbo equalizers can be dramatically
narrowed.\vspace{-0.1cm}
\end{abstract}
\begin{keywords}
Turbo equalization, piecewise linear modelling, hard clustering, soft clustering. \vspace{-0.4cm}
\end{keywords}
\section{Introduction\label{sec:Intro}}\vspace{-0.05in}
Digital communication receivers typically employ a symbol detector to
estimate the transmitted channel symbols and a channel decoder to
decode the error correcting code that was used to protect the
information bits before transmission.  There has been great interest
in enabling interaction between the symbol estimation task and the
channel decoding task, which is often termed ``turbo equalization''
for digital communication over channels with inter-symbol-interference
(ISI).  This interest is due to the dramatic performance gains that
can be obtained with modest complexity\cite{TuechlerC} over performing
these tasks separately. Turbo equalization methods employing
maximum-a-posteriori probability (MAP) detectors demonstrate excellent
bit-error-rate (BER) performance, however their computational
complexity often renders their application impractical
\cite{TuechlerC}.  As an alternative, linear MMSE-based methods offer
comparable performance to MAP-based approaches, with dramatically
reduced complexity \cite{TuechlerC}, compared with the exponential
complexity of the MAP-based approach. However, MMSE-based approaches
still require quadratic computational complexity in the channel length
per output symbol and require adequate channel knowledge or
estimation. To further reduce computational complexity and improve
efficacy over unknown or time-varying channels, ``direct''
LMS-adaptive linear equalizers are often used, employing only linear
complexity \cite{LaotJ} in the regressor vector length, which is often
on the order of the channel delay spread.

While these direct-adaptive methods may reduce computational
complexity and can be shown to converge to their Wiener (MMSE)
solution under stationary environments, they usually deliver inferior
performance compared to linear MMSE-based methods.  A primary reason
for this performance loss is that the Wiener solution is not
time-adaptive, but rather corresponds to the solution of the
``stationarized problem'' where the likelihood information from the
decoder (which is by definition a sample-by-sample probability
distribution over the transmitted data sequence and hence
non-stationary) is replaced by a suitable time-averaged quantity
\cite{LaotJ}. On the other hand, both the linear MMSE and MAP-based
turbo equalizer (TEQ) consider the log-likelihood ratio (LLR) sequence
as time-varying \emph{a priori} statistics over the transmitted
symbols. This LLR information is used to construct the linear MMSE
equalizer, which depends nonlinearly and in a time dependent manner on
the LLR sequence.

In order to reduce the performance gap between LMS-adaptive linear TEQ
and linear MMSE TEQ, we introduce an adaptive approach that can
readily follow the time variation of the soft decision data and
respect the nonlinear dependence of the MMSE symbol estimates on this
LLR sequence while maintaining the low computational complexity of the
LMS-adaptive approach. Specifically, we introduce an
adaptive, piecewise linear equalizer that partitions the space of LLR
vectors from the channel decoder into sets, within which, low
complexity LMS-adaptive TEQs can be used. We use a deterministic
annealing (DA) algorithm \cite{Gersho} for soft clustering the
symbol-by-symbol variances of the transmitted symbols, calculated from
the soft information.  These variances are partitioned into $K$
regions
with a partial membership
according to their assigned
association probabilities \cite{Gersho}. For hard
clustering, the association probabilities are either 1 or 0. In each
cluster, a local linear filter is updated where the contribution to the local update is weighted by the association probabilities \cite{Gersho}.
In addition, we also quantify the mean square error (MSE) of
the approach employing hard clustering and show that it converges
to the MSE of the linear MMSE equalizer as the number of
regions and the data length increase.  In our simulations, we observe that the clustered TEQ significantly improves performance over traditional LMS-adaptive linear equalizers without any significant computational complexity increase.

In Section~\ref{subsec:ateq}, we provide a system description for the communication link under study. The
clustering approach and the corresponding clustered equalization algorithms
are introduced in Section~\ref{sec:algs}. The performance of these algorithms is demonstrated in Section~\ref{sec:simulation}.
We conclude the letter with certain remarks in
Section~\ref{sec:conclusion}.
\section{System Description Under Study\label{subsec:ateq}}
\begin{figure}[tb]
\centerline{\epsfxsize=9.5cm \ \epsfbox{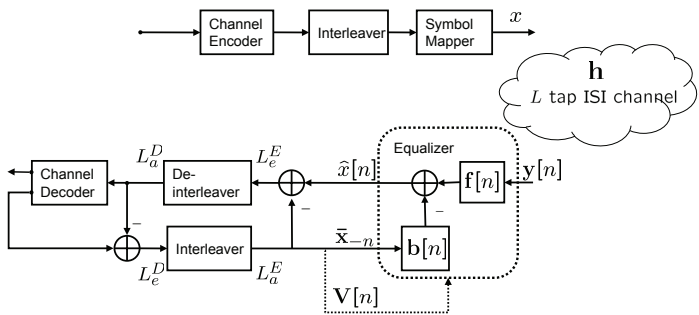}}\vspace{-0.4cm}
   \caption{System block diagram for a bit interleaved coded modulation transmitter and receiver with a linear TEQ}\vspace{-0.7cm}
   \label{fig:blockdiagram}
\end{figure}

\vspace{-0.05in}
We consider the linear turbo equalization system shown in
Fig.~\ref{fig:blockdiagram}.\footnote{All vectors are column vectors
  denoted by lowercase letters and matrices are represented by
  boldface capital letters.  $\vw^H$ is the Hermitian transpose and
  $\|\vw\|$ denotes the $l_2$ norm of $\vw$.  $\mathrm{diag}(\vw)$
  represents the diagonal matrix formed by the elements of $\vw$ along
  the diagonal. For a (random) variable $x$, $\bar{x}=E[x]$. Given $x$
  with a distribution defined from $y$, $E[x : y]$ represents the
  expectation of $x$ with respect to the distribution defined from
  $y$. For a square matrix $\vec{S}$, $\lambda_{\max}(\vec{S})$
  denotes the largest eigenvalue.}  Information bits at the
transmitter are encoded using forward error correction, interleaved in
time, mapped to channel symbols and transmitted through an ISI channel
with impulse response $h_l$, of length $L$, $l=0,\ldots,L-1$ and
additive noise $w[n]$.  The received signal $y[n]$ is given by $y [n]
= \sum_{l=0}^{L-1}{h_l x [n-l]}+ w[n]$, where $h_l$ is assumed time
invariant for notational ease. In \fig{fig:blockdiagram},
the decoder and equalizer pass extrinsic log-likelihood ratio information on the information bits to iteratively improve detection and decoding.
The equalizer produces \emph{a priori} information $L_a^E$ and the decoder computes
the extrinsic information $L_e^D$ which are fed back to the equalizer \cite{TuechlerC}.
For a linear equalizer with a
feedforward filter $\vec{f}$ and feedback filter $\vec{b}$, an estimate of the transmitted
signal can be given by \vspace{-0.05in}
\begin{equation}
    \label{eq:xhat_MMSE}
   \hat{x}[n] = \vec{f}^H[n] \by[n] - \vec{b}^H[n] \bar{\bx}_{-n}[n],
\end{equation}
where $\vec{y} [n] = [y [n-N_2],\cdots, y
[n+N_1] ]^T$, $\bar{\bx}_{-n}[n] = [\bar{x}
[n-N_2-L+1], \cdots, \bar{x} [n-1], \bar{x} [n+1], \cdots, \bar{x}
[n+N_1] ]^T$. The mean symbol values are calculated using the \emph{a
priori} information $L_{a}^E$ provided by the SISO decoder, i.e.,
$\bar{x} [n] = E [x [n] : \{L_{a}^E\}]$ and $E [|x [n]|^2 : \{L_{a}^E\}]=1$
\cite{TuechlerC}, where we assumed BPSK signaling for notational simplicity. If a linear MMSE equalizer is used in \eqref{eq:xhat_MMSE}, we get \vspace{-0.05in} 
{\small\begin{equation}
\bff[n] = ( \bH_{-0} \bV[n] \bH_{-0}^H+ \bs \bs^H+  \sigma_w^2 \bI)^{-1} \bs ,
\vec{b}[n] =  \bH_{-0}^H \bff[n], \label{eq:sol_MMSE2_i}
\end{equation}}
where $\vec{H}$ is the channel convolution matrix of size $N \times (N
+L-1)$, $\bs$ is the $(N_2 + L)$th column of $\bH$,
$\bH_{-0}$ is the matrix where the $(N_2 + L)$th
column of $\bH$ is eliminated, $\bV [n] = \text{diag}([v
  [n-N_2-L+1], \cdots, v [n-1], v [n+1], \cdots, v[n+N_1]
  ])$, $v [n] = E[ |x [n]|^2 : L_{a}^E ]
- |\bar{x} [n]|^2$ and $\sigma_w^2$ is the additive noise variance assuming
fixed transmit signal power of $1$.
\begin{remark}
The linear MMSE equalizer in \eqref{eq:sol_MMSE2_i} is time varying
due to the symbol-by-symbol variation of the soft input variance, $\bV[n]$, even
if $h_l$ is time invariant. The linear MMSE equalizer is a nonlinear
function of $\bV[n]$. If $h_l$ is also time varying, then
\eqref{eq:sol_MMSE2_i} could be readily updated by including this time
variation.
\end{remark}

Unlike the linear MMSE equalizer, ``direct'' adaptive linear TEQs use
adaptive updates (e.g. using LMS or RLS), for direct estimation of the
transmitted symbols by processing the received signal and LLR
information without the need for channel estimation \cite{LaotJ}. In
general, these approaches use only the mean vector $\bar{\bx}_{-n}$ as
feedback, i.e., soft decision data are not considered as \emph{a
  priori} probabilities, where each component of $\bar{x}$ is taken as
a random variable with zero mean and variance $\sigma_{\bar{x}}^2$.
As an example, if one uses the NLMS direct adaptive linear equalizer, we have the update\vspace{-0.05in}
\begin{align}
& e[n] =   \tilde{x} [n] - \bw^H [n] \bu [n], \nn\\
& \bw[n+1] =\bw [n] + \mu e^* [n] {\bu [n]}/{\| \bu [n]\|^2}, \nn
\end{align}
where $\bw [n+1] = [\bff^H [n+1] \; -\bb^H [n+1]
  ]^H$, $\bu [n] = [ \by^H [n] \; \bar{\bx}_{-n}^H
  ]^H$, $\mu$ is the step size and $\tilde{x}[n]$ is equal to the mean $\bar{x}[n]$.
Under this stationarity assumption on $\bar{x}$ and LLRs, the feedforward filter
using $\bar{\bx}_{-n}$ converges to the MSE optimal Wiener (stationary MMSE) solution \vspace{-0.05in}
\begin{equation}
    \label{fnNLMS}
  \bff = ((1 - \sigma_{\bar{x}}^2)\bH_{-0}\bH_{-0}^H +\bs\bs^H+\sigma_w^2 \bI)^{-1}\bs 
\end{equation}
and $\bb =\bH_{-0}^H \bff$, assuming zero variance at
convergence \cite{TuechlerC}.
The resulting filter in \eqref{fnNLMS} at convergence is time invariant
and is identical to \eqref{eq:sol_MMSE2_i} with time
averaged soft information \cite{TuechlerC}.
The linear MMSE in \eqref{eq:sol_MMSE2_i} requires $O((N+L)^2)$ computations per output, however,
\eqref{fnNLMS} requires only $O(N+L)$. Since \eqref{fnNLMS} is not time varying and implicitly assumes that the soft information is stationary, there is a large performance gap between linear MMSE in \eqref{eq:sol_MMSE2_i} and \eqref{fnNLMS} \cite{TuechlerC}.
We seek to reduce this performance gap between the
direct adaptive methods with respect to the linear MMSE
approach, by capturing the nonlinear dependence of the MMSE solution on the soft-information, without capturing the associated computational complexity of
\eqref{eq:sol_MMSE2_i}.
\vspace{-0.3cm}\section{Adaptive Turbo Equalization Using Hard or Soft
  Clustered Linear Models\label{sec:algs}}\vspace{-0.05in} We propose
to use adaptive local linear filters to model the nonlinear dependence
of the linear MMSE equalizer on the variance computed from the soft
information generated by the SISO decoder in
\eqref{eq:sol_MMSE2_i}. We do this by partitioning the space of
variances in \eqref{eq:sol_MMSE2_i} into a set of regions within each
of which a single direct adaptive linear filter is used. As a result,
we can retain the computational efficiency of the direct adaptive
methods, while capturing the nonlinear dependence (and hence
sample-by-sample variation) of the MMSE optimal TEQ.
\vspace{-0.3cm}\subsection{Adaptive Nonlinear Turbo Equalization Based on Hard Clustering\label{subsec:hard}}\vspace{-0.05cm}
Suppose a hard clustering algorithm is applied to $\{\vec{v}[n]\}_{n \geq 1}$ after
the first turbo iteration to yield $K$ regions $\cR_k$, with
the corresponding centroids $\tilde{\vec{v}}_k$, $k=1,\ldots,K$. Here, $\vv[n]$
is the vector formed by the diagonal entries of $\bV [n]$.
As an example,  one might use the $K$-means algorithm (LBG
VQ)\cite{Gersho}.
In the LBG VQ algorithm, the centroids and the corresponding regions are determined as
$
\tilde{\vv}_k \defi \sum_{n, \vv[n]\in \cR_k} \vv[n]/\left(\sum_{n, \vv[n] \in \cR_k}1\right), $
and $\cR_k \defi \{ \vv:\|\vv-\tilde{\vv}_k\| \leq \|\vv-\tilde{\vv}_i\|, i = 1,\ldots, K, i \neq k \}$,
where the regions $\cR_k$ are selected using a greedy algorithm
\cite{Gersho}.  After the regions are constructed using the VQ
algorithm, the corresponding filters in each region are trained with
an appropriate direct adaptive method, and the estimate of $x[n]$ at each
time $n$ is computed as $\hat{x}[n] = \hat{x}_i[n]$ if $i = \arg\min_k
\|\vv[n]-\tilde{\vv}_k\|$. For the adaptive algorithms to
converge in each of these regions, we put a constraint on the cluster-size such that each cluster contains at least $N_{\min}$ (the
minimum required data length for suitable convergence)
elements and the quantization level is equal to or less than that of
the original LBG VQ. At each time $n$, the received data is assigned
to one of the regions and used in an adaptive algorithm to train a
locally linear direct adaptive equalizer. For a locally NLMS direct adaptive linear
equalizer, we have the update \vspace{-0.05in}
\begin{align}
& e_k [n] =   \tilde{x} [n] - \bw_k^H [n] \bu [n], \label{eq:wnLMS1} \\
& \bw_k [n+1] =\bw_k [n] + \mu e_k^* [n] \bu [n]/(\| \bu [n]\|^2), ~\bv [n] \in \cR_k, \nn \\
& \bw_i [n+1] =\bw_i [n], i=1,\ldots,K, i \neq k, \label{eq:wnLMS} \\
& \hat{x}[n] = \hat{x}_k[n], \nn
\end{align}
where $\bw_k [n+1] = [\bff_k^H [n+1] \; -\bb_k^H [n+1]
  ]^H$, $\bu [n] = [ \by^H [n] \; \bar{\bx}_{-n}^H
  ]^H$, and $\tilde{x}[n]$ in
\eqref{eq:wnLMS1} is equal to either the hard quantized $\hat{x}[n]$
or the mean $\bar{x}[n]$ in decision directed (DD) mode.
An algorithm description is given in Table~\ref{tab:TEQ_hard}.
Here, $L_T$ and $L_D$ are the length of training data and transmit
data. During training period, perfect knowledge for 
the transmitted data $x[n]$ is available, so the $K$ adaptive filters can use weighted
training symbols as input to the feedback filters in order to enable
the filters to converge to a function of the quantized soft input
variance. The weight matrices are selected as
$(\vec{I}-\tilde{\vec{V}}_{k^{(i)}})$ at the $i$th turbo iteration.
\begin{table}
{\fontsize{6}{6}\selectfont
\begin{center}
\parbox{1\linewidth}{
\centering
\caption{Pseudocode for adaptive TEQ via hard clustering} \label{tab:TEQ_hard}\vspace{-0.3cm}
\begin{tabular}{l} \hline \vspace{0.1cm}
Set $N_{\min}$. $K_1=\lfloor {L_D}/{N_{\min}} \rfloor$, \hfill (line A)\\
$i = 1$, $\%$ First turbo iteration \\
for $k=1:K+1$; $\vec{w}_{k^{(1)}} = \vec{0}$, endfor \\
for $n=1:L_T$; \\
\hspace{0.2in} $e[n] =x[n]-\vec{w}_{(k+1)^{(1)}}^H[n]\vu[n]$,\\
\hspace{0.2in} $\vec{w}_{(k+1)^{(1)}}[n+1] = \vec{w}_{(k+1)^{(1)}}[n]+ \mu e^*[n] \vu[n]$, endfor \\
for $n=L_T+1:L_T+L_D$; \\
\hspace{0.2in} $e[n] =\tilde{x}[n]-\vec{w}_{(k+1)^{(1)}}^H[n]\vu[n]$,\\
\hspace{0.2in} $\vec{w}_{(k+1)^{(1)}}[n+1] = \vec{w}_{(k+1)^{(1)}}[n]+ \mu e^*[n] \vu[n]$, endfor \\
for $i=2,\ldots,$ $\%$ turbo iterations, \\
{\bf Perform hard clustering, based on modified LBG algorithm.} \hfill (line B)\\
\hspace{0.2in} Outputs: $K_i =K$, $\vec{V}_{k^{(i)}}=\vec{V}_k$, \hfill (line C)\\
\hspace{0.2in} for $k^{(i)} =1:K_i$, {$\%$ Filter initialization}\\
\hspace{0.3in} if $i==2$; $\bff_{k^{(i)}} [1] = \bff_{K_1+1} [L_T+L_D]$, \\
\hspace{0.3in} else $k^* = \argmin{k^{(i-1)}}{\|\tilde{\bV}_{k^{(i)}} - \tilde{\bV}_{k^{(i-1)}}\|^2}$,\\
\hspace{0.3in} $k^{(i-1)} = 1, \ldots, K_{i-1}$, $\bff_{k^{(i)}} [1] = \bff_{k^*} [L_T +L_D]$, \\
\hspace{0.3in} $\bb_{k^{(i)}}[1] = \bb_{k^*}[L_T +L_D]$,  endfor \\
\hspace{0.2in} for $n=1: L_T$, $\%$ Training period. \\
\hspace{0.3in} $\vw_k[n+1] = \vw_k[n]+\mu_k e_k^*[n](\bI - \tilde{\bV}_{k^{(i)}})^{1/2} \vec{u}[n]$, endfor \\
\hspace{0.2in} for $n=L_T+1: L_T+L_D$;\\
\hspace{0.3in} $k^* = \argmin{k^{(i-1)}}{\|\tilde{\bV}_{k^{(i)}} - {\bV} [n]\|^2}$ \hfill (line D)\\
\hspace{0.3in} $\bw_{k^*} [n+1] =\bw_{k^*} [n] + \mu_{k} [n] e_k^* [n] {\bu [n]}/{\| \bu [n]\|^2}$, \hfill (line E)\\
\hspace{0.3in} $\mu_{k} [n] = \left\{\begin{array}{l}{\mu \text{ for } k=k^*}  \\ {0}\\ \end{array} \right.$, $\hat{x}[n] =\bw_{k^*}^T[n] \bu[n]$ endfor \hfill (line F)\\
	{\textbf{Go to the Clustering step}: Until desired turbo iterations or error rate} \\ \hline	\vspace{-1.2cm}
\end{tabular}}
\end{center}}
\end{table}
Note that the complexity of the locally linear adaptive filters are
higher than direct equalization due to the clustering step.
Since the clustering is only performed at the
start of each iteration with $O(N+L-1)$ complexity per data
symbol, the equalization complexity is effectively unchanged per output
symbol. If the regions are dense enough such that
$\vec{v}[n] \approx \tilde{\vec{v}}_k$ for all regions, then the
adaptive filter in the $k$th region converges to $
\bff_k = ( \bH_{-0} \tilde{\bV}_k \bH_{-0}^H+
\bs \bs^H+ \sigma_w^2 \bI)^{-1}\bs,$
$\tilde{\vec{V}}_k = \mathrm{diag}(\tilde{\vec{v}}_k)$, assuming
zero variance at convergence.
The difference between the MSE of the converged filter $\bff_k$ and the MSE of the linear MMSE equalizer is given as \cite{TuechlerC}
\begin{equation}
\label{MSEdiff}
\bff_k^H \bH_{-0} (\bV[n] - \tilde{\bV}_k) \bH_{-0}^H \bff_k + (1 - \bff_k^H \bs) - (1-\bff_n^H \bs ).\vspace{-0.05cm}
\end{equation}
By defining {$\bA= ( \bH_{-0} \tilde{\bV} \bH_{-0}^H+ \bs
\bs^H+ \sigma_w^2 \bI)$}, { $\bB  = \bA + \bH_{-0} \bE
\bH_{-0}^H $} and {$\bE =
\bV-\tilde{\bV}$}, the difference \eref{MSEdiff} yields \vspace{-0.05in}
\begin{align}
& \bs^H \bA^{-1} \bH_{-0} \bE \bH_{-0}^H \bA^{-1} \bs + \bs^H ( \bB^{-1}- \bA^{-1})\bs  \nn\\
&=\bs^H \bA^{-1} \bH_{-0} \bE \bH_{-0}^H \bB^{-1} \bH_{-0} \bE \bH_{-0}^H \bA^{-1} \bs \label{bound1:1} \\
&\leq \lambda_{\mathrm{max}}(\bH_{-0} \bE \bH_{-0}^H \bB^{-1} \bH_{-0} \bE \bH_{-0}^H ) \bs^H \bA^{-2} \bs \label{bound1:2} \\
&\leq e_{\mathrm{max}}^2 \lambda_{\mathrm{max}}^2 (\bH_{-0} \bH_{-0}^H ){\lambda_{\mathrm{min}} (\bB )} \bs^H \bA^{-2} \bs\nn,
\end{align}
where $e_{\mathrm{max}}$ is the maximum element of the error
diagonal matrix $\bE$. Here, \eqref{bound1:1} follows from $(
\bB^{-1} -\bC^{-1} ) =\bB^{-1} ( \bC -\bB)\bC^{-1}$,
\eqref{bound1:2} follows from $\mathrm{tr} (\bC \bD) = \mathrm{tr}(\bD
\bC)$ and $\mathrm{tr}(\bC \bD) \leq \lambda_{\mathrm{max}} (\bC)
\mathrm{tr}(\bD)$, and the last line follows from $
\lambda_{\mathrm{max}} (\bC \bD) \leq \lambda_{\mathrm{max}} (\bC)
\lambda_{\mathrm{max}} (\bD)$. Since $\lambda_{\mathrm{min}} (\bB
) \geq \sigma_w^2$ and $\lambda_{\mathrm{max}} (\bH_{-0}
\bH_{-0}^H ) \leq \lambda_{\mathrm{max}} (\bH \bH^H )
\leq (\sum_{m}{|h_m|})$ for the Toeplitz matrix $\bH$, the MSE difference in \eqref{MSEdiff} is bounded by
$C e_{\mathrm{max}}^2$ for some $C < \infty$. Hence, the MSE of the
hard clustered linear equalizer converges to the MSE of the linear
MMSE equalizer as the number of the regions increase provided there is
enough data for training.
\subsection{Adaptive Nonlinear Turbo Equalization Based on Soft Clustering\label{subsec:soft}}\vspace{-0.05cm}
Suppose the deterministic annealing (DA) algorithm described in
Table \ref{tab:DA} is used for soft clustering 
\cite{Gersho} on $\{\vec{v}[n]\}_{n \geq 1}$ after the first turbo
iteration, to give $K$ clusters with the corresponding centroids
$\tilde{\vec{v}}_k$ and association probabilities
$P(v[n]|\tilde{\vec{v}}_k)$, $k=1,\ldots,K$. Then, at each time
$n$, the vector $\vec{v}[n]$ can be partially assigned to all
$K$ regions using conditional probabilities yielding the update\vspace{-0.05in}
\begin{align}
& e_k[n] = \tilde{x}[n]-\vec{w}_k^H[n]\vec{u}[n], \nn \\
& \bw_k [n+1] =\bw_k [n] + \mu_{k} [n] e_k^* [n] \bu [n]/(\| \bu [n]\|^2), \label{eq:wnLMSsoft}\\
& \mu_{k} [n] = \mu P\left(\bv[n] \vert \tilde{\bv}_k \right), \label{eq:wnLMSsoft2}
\end{align}
where $\bw_k [n+1] = [\bff_k^H [n+1] \;\; -\bb_k^H [n+1] ]^H$, $\bu
[n] = [ \by^H [n] \;\; \bar{\bx}_{-n}^H ]^H$ and $\mu_{k}[n]$ is the
fractional step size. To generate the final output, outputs of $K$
linear filters can be combined by either using another adaptive
algorithm \cite{Kozat} or other combination methods \cite{Gersho}.
We use the method in \cite{Kozat} as follows. At each time $n$,
we construct $\vec{y}[n] = [\hat{x}_1[n],\ldots,\hat{x}_K[n]]^T$
and produce the final output and update the weight vectors as \vspace{-0.05in}
\begin{align}
& \hat{x}[n] =\vec{w}^T[n] \vec{y}[n], \label{eq:comb1}\\
& e[n]  = \tilde{x}[n] - \vec{w}^T[n]\vec{y}[n], \label{eq:comb2} \\
& \bw [n+1]  =\bw [n] + \mu [n] e^* [n] {\vec{y} [n]}/{\| \vec{y} [n]\|^2}, \label{eq:comb}
\end{align}
and $\mu$ is a learning rate for this combining step. An update as in \eqref{eq:comb} can
provide improved steady-state MSE and convergence speed exceeding that of any of
the constituent filters, i.e., $\hat{x}_k[n]$, $k=1,\ldots,K$, under
certain conditions \cite{Kozat}.

The algorithm description is the same as in Table~\ref{tab:TEQ_hard},
except that line A is removed and $K_1$ is set to $K_{\mathrm{max}}$,
and soft clustering \cite{Gersho} is used in line B.
In line C, we add an $L_D \times K_i$ probability matrix corresponding
to $P \left( \bv [n] \vert \tilde{\bv}_k \right)$ to the outputs.
Line D and E are removed, \eqref{eq:wnLMSsoft} and \eqref{eq:wnLMSsoft2}
for all $k$ are used instead. Line F is also removed and replaced by
\eqref{eq:comb1}, \eqref{eq:comb2} and \eqref{eq:comb}, respectively.
\begin{table}
{\fontsize{6}{6}\selectfont
\caption{Soft Clustering based on Deterministic Annealing} \label{tab:DA} \vspace{-0.5cm}
\begin{center}
\parbox{1\linewidth}{
\centering
\begin{tabular}{l} \hline
$\%$ Set the maximum number of code vectors, the maximum number of iterations\\
$\%$ and a minimum temperature, i.e., $K_{\mathrm{max}}$, $I_{\mathrm{max}}$ and $T_{\mathrm{min}}$. \\
$K=1$, $\tilde{\bv}_1 = \frac{1}{N}\sum_{n} \bv [n]$ and $P(\tilde{\bv}_1) = 1$. \\
$T=T_0$ $\%$ An initial temperature, $T_0$, should be larger than $\lambda_{\mathrm{max}} \left(\mathrm{cov} \left(\bv, \bv \right) \right)$.\\
for $n=1:N$; \hspace{0.1cm} $P(\tilde{\bv}_k \vert \bv [n])=\frac{1}{N}$ endfor\\
$D = \frac{1}{N} \sum_n {d(\bv [n],\tilde{v}_1)}$\\
\hline
if $T \geq T_{\min}$;\\
\hspace{0.1in} $T=aT$ for $a<1$ $\%$ Cooling Step\\
\hspace{0.1in} if $K \leq K_{\mathrm{max}}$; $j = 0$.\\
\hspace{0.2in} for $k=1:K$;\\
\hspace{0.3in} if $T >Tc_k$; $\%$ Split the $k$th cluster with slight perturbation\\
\hspace{0.3in} elseif $j=j+1$ endfor\\
\hspace{0.2in} if $j == K$; finish DA.\\
\hspace{0.1in} elseif; finish DA\\
\hspace{0.1in}elseif; finish DA\\
\hline
$i=1$\\
while converged or $i<I_{\mathrm{max}}$;\\
\hspace{0.1in}for $k=1:K$; \\
\hspace{0.2in}for $n=1:N$;\\
\hspace{0.3in}  $P(\tilde{\bv}_k \vert \bv [n]) = P(\tilde{\bv}_k) exp(- \frac{\|\bv [n] - \tilde{\bv}_k \|^2}{T})/\sum_k {P(\tilde{\bv}_k) exp(- \frac{\|\bv [n] - \tilde{\bv}_k \|^2}{T})} $ \\
\hspace{0.2in}$P(\tilde{\bv}_k) = \sum_n P(\bv [n]) P(\tilde{\bv}_k \vert \bv [n])$, $\tilde{\bv}_k = \frac{\sum_{n}{\bv [n] P \left(\tilde{\bv}_k \vert \bv [n]\right) P\left(\bv [n]\right)}}{P \left(\tilde{\bv}_k \right)}$\\
\hspace{0.1in}endfor $\%$ calculate distortion and check convergence\\
endwhile $\%$ Go to Cooling Step\\ \hline
\end{tabular}
\vspace{-0.9cm}}
\end{center}}
\end{table}
\vspace{-0.4cm}\section{Simulation
  Results\label{sec:simulation}}
\begin{figure*}[t!]
\vspace{-0.15cm}
\hfill
\begin{minipage}[t]{.32\textwidth}
  \begin{center}
  \epsfxsize=2.6in {\epsfbox{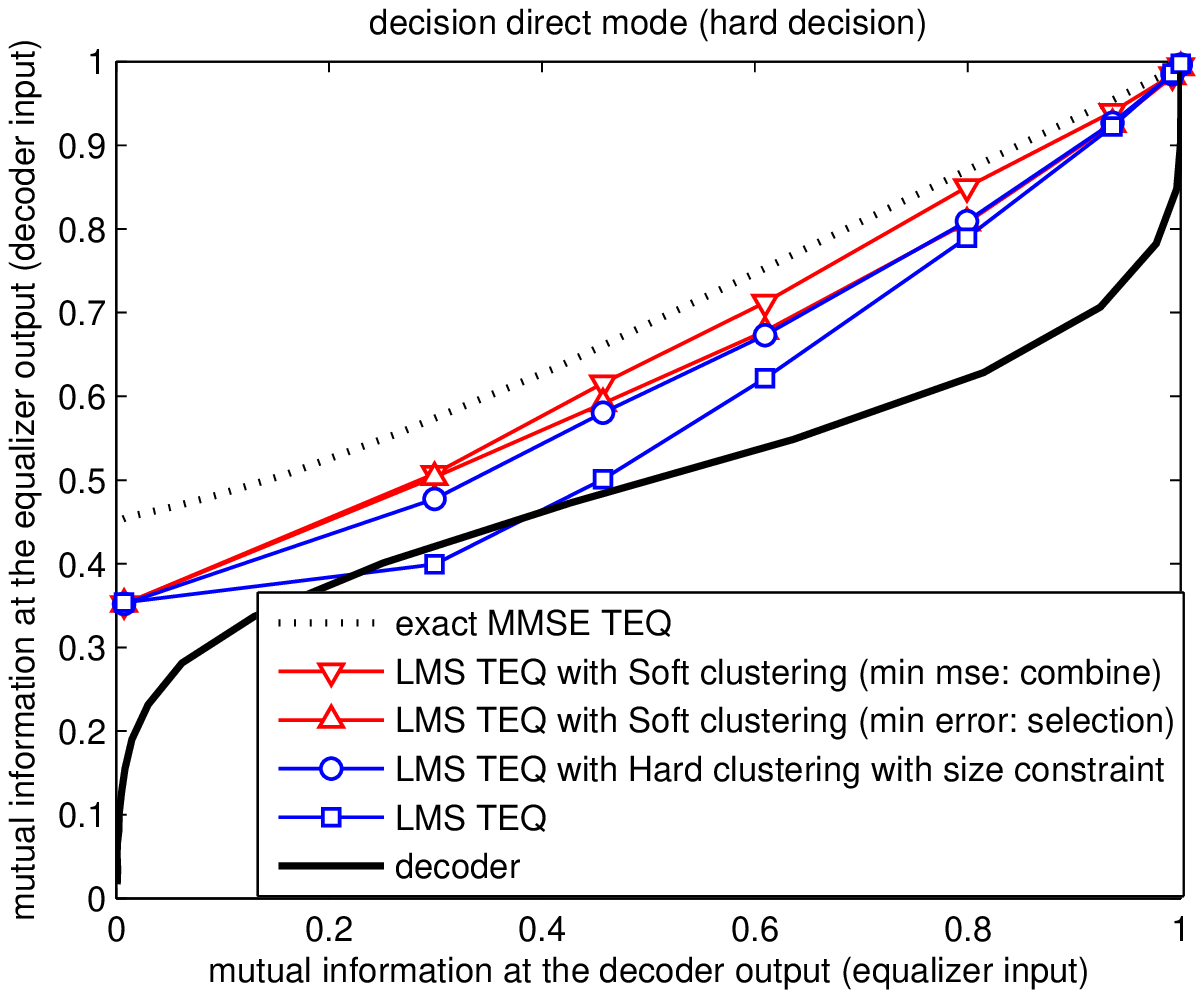}}\vspace{-0.4cm}
  \caption{EXIT chart comparison in DD mode. ($Eb/N0 = 10dB$)}\vspace{-0.2cm}
   \label{fig:EXITDD}
   \end{center}
\end{minipage}
\hfill
\begin{minipage}[t]{.32\textwidth}
  \begin{center}
     \epsfxsize=2.6in {\epsfbox{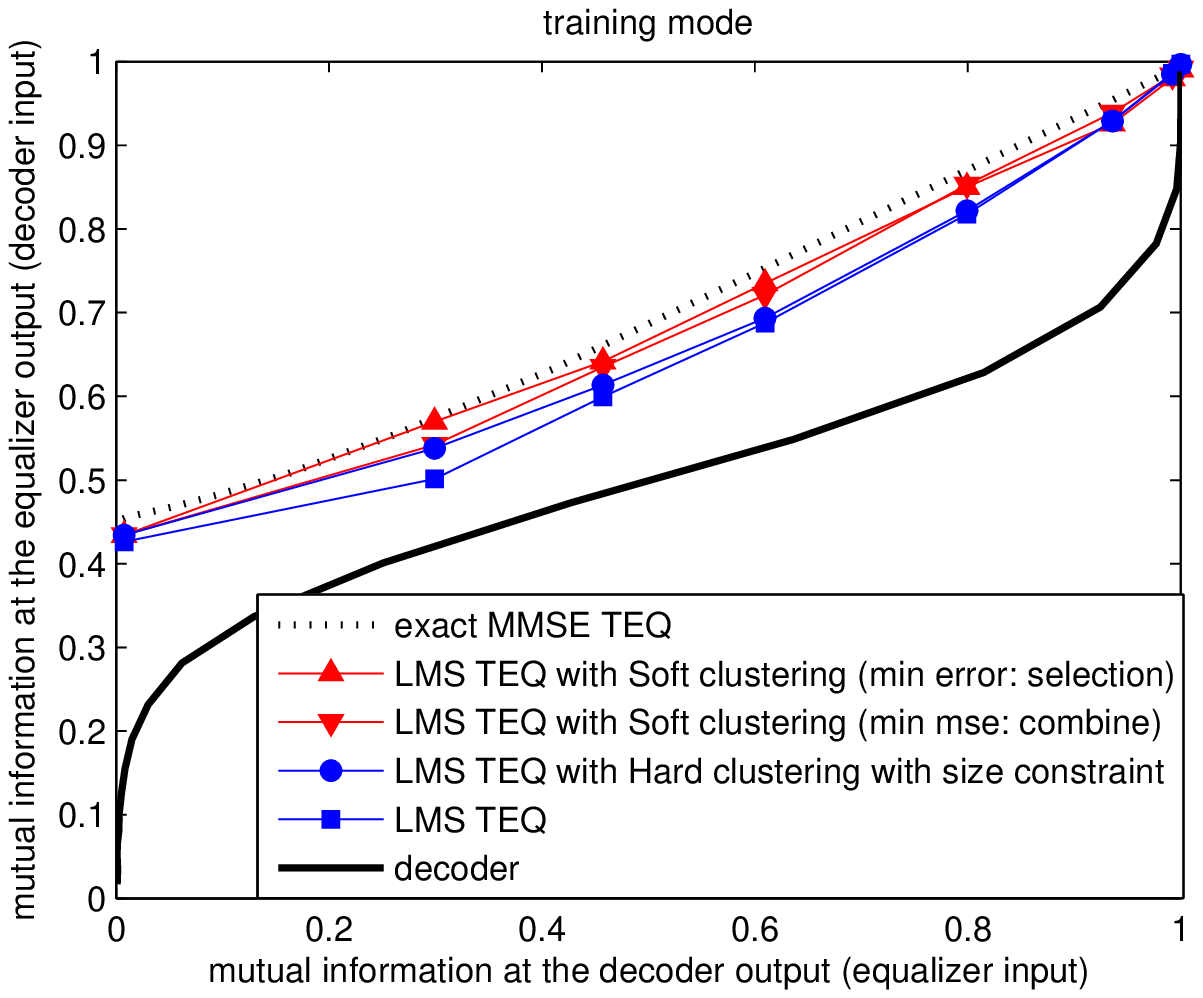}}\vspace{-0.4cm}
     \caption{EXIT chart comparison in training mode. ($Eb/N0 = 10dB$)}\vspace{-0.2cm}
     \label{fig:EXITtrain}
  \end{center}
\end{minipage}
\hfill
\begin{minipage}[t]{.32\textwidth}
  \begin{center}
     \epsfxsize=2.6in {\epsfbox{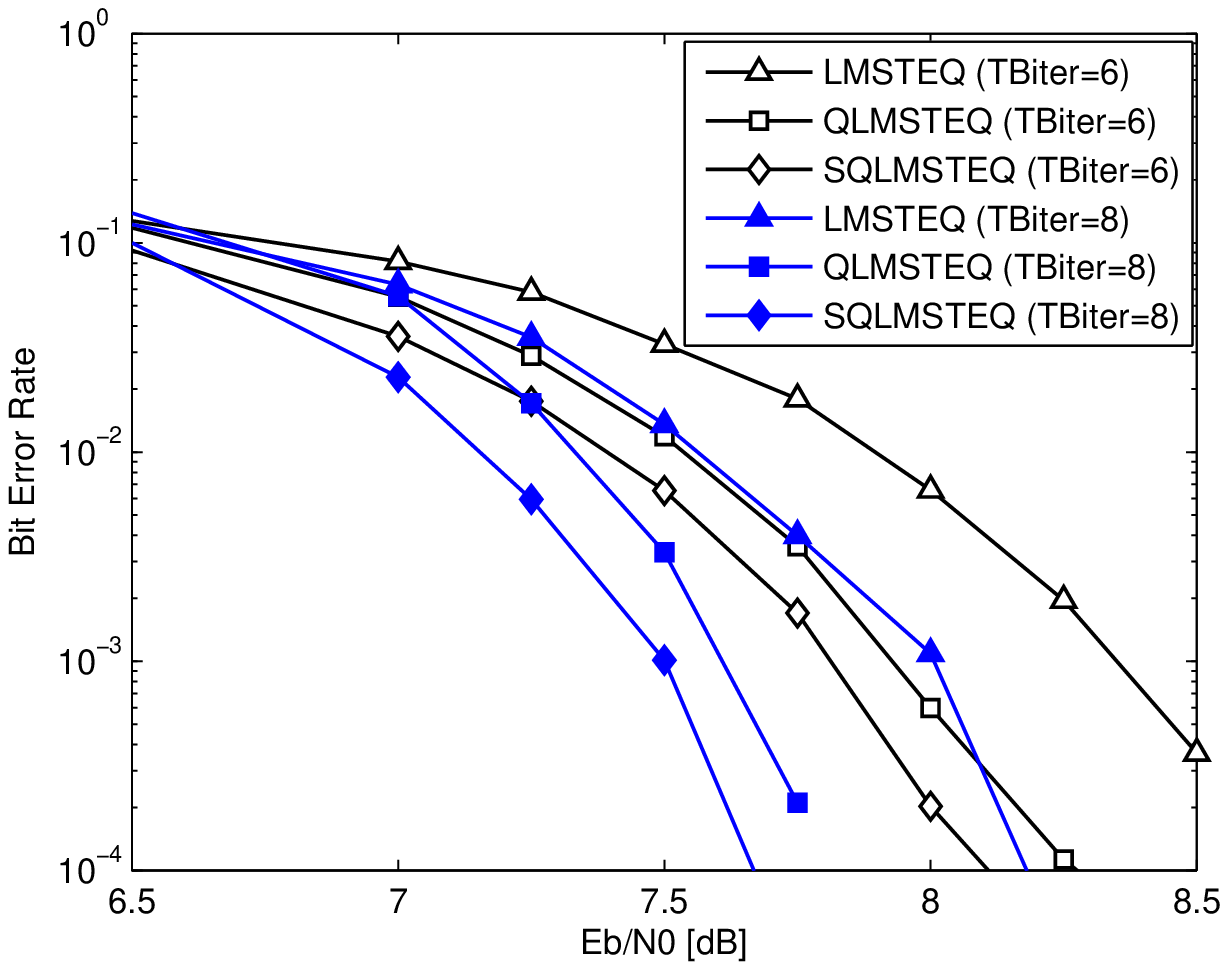}}\vspace{-0.4cm}
     \caption{BER comparison in DD mode, where soft decision value from the decoder is used and TBiter is the turbo iteration count. }\vspace{-0.2cm}
   \label{fig:ber}
  \end{center}
\end{minipage}
\hfill
\end{figure*}

\vspace{-0.05in} Throughout the
simulations, a time invariant ISI channel given by $h_l=[0.227, 0.46,
  0.688, 0.46, 0.227]$ is used. We use rate $1/2$ convolutional code
with constraint length $3$, random interleaving and BPSK signaling. We
choose $L_T=1024$, $L_D=4096$, $N_{\mathrm{min}} = 500$ and
$K_{\mathrm{max}} = 8$. Each NLMS filter has a length 15 feedforward
and length 19 feedback filter ($N_1=9,N_2=5$) where $\mu=0.03$. For an
NLMS filter with soft clustered TEQ, the filter length is less than
$K_{\mathrm{max}}$ and $\mu=0.1$.  Fig. \ref{fig:EXITDD} and
Fig. \ref{fig:EXITtrain} show EXIT charts for a conventional NLMS TEQ
\cite{LaotJ} (LMSTEQ), the switched NLMS TEQ based on hard clustering
with restriction on the number of data samples in each cluster
(QLMSTEQ) and an NLMS TEQ based on soft clustering (SQLMSTEQ). In
\fig{fig:EXITDD}, hard decision data are used to learn the NLMS
filter, while in \fig{fig:EXITtrain} the transmitted signals are used
during data the data transmission period. In \fig{fig:ber}, we provide
the corresponding BERs.
For the soft clustering based NLMS TEQ, the final output is given by
either adaptively combining to minimize combined MSE with another NLMS
filtering as given in Section \ref{subsec:soft} or selecting one of
the outputs to minimize instantaneous residual error after
filtering.

In all simulations, adaptive TEQs based on soft clustering
showed significantly better performance to hard clustered
adaptive TEQ and direct adaptive TEQ. In Fig.~\ref{fig:EXITDD},
(i.e., in the DD mode with hard decision data), the adaptive
combination of adaptive filters showed better performance
than selecting a single filter, since the combination method can mitigate the worst-case 
selection \cite{Kozat}. However, in a dynamically
changing feature domain, combining the outputs of the constituent filters
in MSE can loose the benefit from the local linear models \cite{Kozat}.
As shown in \fig{fig:EXITtrain}, selecting one filter among $K$ filters shows
better performance than the combination of the filters.  As
discussed in Section \ref{sec:algs}, the DD-NLMS TEQ can achieve 
``ideal'' performance, i.e. time-average MMSE TEQ, as the decision
data becomes more reliable. However, there is still a mutual information
gap between the exact MMSE TEQ and the NLMS adaptive TEQ. As an example, the
NLMS TEQ in \fig{fig:EXITDD} cannot converge to its ideal performance
if the tunnel between the transfer function of equalizer and that of
the decoder is closed. This point can be identified by measuring the signal to noise
ratio (SNR) threshold. If the SNR is higher than the SNR threshold, turbo
equalization can converge to near error-free operation. Otherwise, turbo equalization stalls, and fails to improve after a few iterations. The $\frac{Eb}{N0}$s corresponding to the SNR thresholds by equalization algorithm are given in Table
\ref{tab:snrth}. Adaptive nonlinear TEQs based on soft clustering yielded
$0.5dB \frac{Eb}{N0}$ gain in SNR threshold compared to adaptive
nonlinear TEQ based on hard clustering and about $1dB
\frac{Eb}{N0}$ gain compared to the conventional adaptive linear TEQ.
\bet {\fontsize{6}{6}\selectfont
\caption{SNR thresholds in $dB$ of several algorithms}  \label{tab:snrth}
\begin{center}
\vspace{-0.6cm}
\begin{tabular}{| c || c | c |} \hline
	{mode}	 	 & decision directed  & training  \\ \hline \hline
    {original NLMS TEQ}  & 10.9 & 6.0 \\  \hline
    {NLMS TEQ w/ hard clustering}  & 6.5 & 5.5\\  \hline
    {NLMS TEQ w/ soft clustering (combine)} & 5.3 & 5.0 \\  \hline
    {NLMS TEQ w/ soft clustering (selection)} & 5.9 & 4.8 \\  \hline
\end{tabular}
\vspace{-0.9cm}
\end{center}
}
\eet
\vspace{-0.2cm}\section{Conclusion\label{sec:conclusion} }\vspace{-0.15cm}
We introduced adaptive locally linear filters
based on hard and soft clustering to model the nonlinear dependency of
the linear MMSE turbo-equalizer on soft information from the decoder. The adaptive equalizers have computational complexity on the order of an
ordinary direct adaptive linear equalizer.  The local adaptive filters
are updated either based on their associated region using hard clustering or
fractionally based on association probabilities in soft clustering.
Through simulations, the superiority of the proposed algorithms are
demonstrated.\vspace{-0.02cm}

\def\ninept{\def\baselinestretch{0.6}}
\ninept
\bibliographystyle{IEEEtran}
{\small \vspace{-0.3cm}\bibliography{IEEEabrv,DAturbo}\vspace{-0.1cm}}
\end{document}